# ENERGY PHASE CORRELATION AND PULSE DYNAMICS IN SHORT BUNCH HIGH GAIN FELs

G. Dattoli, L. Giannessi, P. L. Ottaviani and S. Pagnutti

## ABSTRACT

We analyze the dynamics of Free Electron Laser (FEL) devices, operating with a bunched beam exhibiting a longitudinal phase space correlation. We show that the presence of an energy-position correlation term is responsible for very interesting effects like an enhancement of the peak output power, a shortening of the laser pulses and an increase of the non linearly generated harmonic intensities. We conjecture that the mechanism is due to a kind of energy tapering effect associated with the correlation. We discuss the difference of the dynamics with respect to an ordinary undulator tapering and the relative advantages.

## 1 Introduction

The FEL high gain dynamics is strongly affected by the characteristics of the electron beam phase space distribution. The effects of transverse phase space have been thoroughly investigated [1]. More recently, new concepts, associated with the slice phase space distribution, have emerged as a consequence of the peculiar properties of the FEL SASE dynamics [2], characterized by the so called coherence length, due to the slippage mechanism [3].

The FEL radiation, produced by a single electron bunch, slips, indeed, over the bunch itself, thus creating a kind of longitudinal mode-locking [4], responsible for a "local" coherence, due to the fact that radiation spans, during the interaction, over a small portion of the bunch only. Different, uncorrelated "local mode locked structures", distributed all over the bunch, may interfere destructively during the interaction, thus giving rise to the spiking behavior characterizing the SASE FEL radiation [5]. This is indeed one of the main problems, which may hamper the use of the FEL radiation, for applications requiring a good deal of coherence.

It is evident that an electron bunch, with a length comparable to the coherence length, would provide the natural solution to this problem.

In this paper we develop a systematic investigation of the characteristics of the FEL dynamics, with bunch length comparable to the coherence length and exhibiting also an energy position correlation [6].

We consider therefore a FEL driven by an electron bunch characterized by a longitudinal distribution of the type

$$f(z,\varepsilon) = \frac{1}{2\pi\Sigma_\varepsilon}\exp\left(-\frac{\gamma_\varepsilon z^2 + 2\alpha_\varepsilon z\varepsilon + \beta_\varepsilon \varepsilon^2}{2\Sigma_\varepsilon}\right),$$
$$\varepsilon = \frac{E-E_0}{E_0} \qquad (1).$$

where $\varepsilon$ is the relative energy and $\Sigma_\varepsilon$ is the longitudinal emittance, which, along with the Twiss parameters, define the bunch length and the relative energy spread as

$$\sigma_\varepsilon = \sqrt{\gamma_\varepsilon \Sigma_\varepsilon},$$
$$\sigma_z = \sqrt{\beta_\varepsilon \Sigma_\varepsilon}, \qquad (2).$$

Furthermore the normalization of the distribution reported in eq. (1) requires that

$$\beta_\varepsilon \gamma_\varepsilon - \alpha_\varepsilon^2 = 1 \qquad (3).$$

The parameter $\alpha_\varepsilon$ accounts for the energy position correlation and the FEL operation, with an electron bunch having such a correlation, displays a very interesting dynamical behavior.

Before entering more specific aspects we note that electrons, exhibiting an energy correlation along the bunch, are characterized by an energy which depends on the position $z_b$ inside the bunch specified by the relation

$$E(z_b) = E_0\left(1 - \frac{\alpha_\varepsilon}{\beta_\varepsilon} z_b\right) \qquad (4)$$

If $\alpha_\varepsilon$ is negative the tail of the bunch will have less energy than the head. The electrons of the bunch are therefore radiating at different wavelengths, which becomes longer at the tail. The situation is reminiscent of a kind of energy tapering which may have consequence on the output radiation characteristics and in particular on the power, which should be larger than that obtained in the uncorrelated case.

This hypothesis is confirmed by the numerical computation, and in Fig. 1 we have reported the maximum power vs. $\alpha_\varepsilon$ for FEL parameters analogous to those presently employed at SPARC (undulator period $\lambda_u \cong 2.8\,cm$, parameter strength $K \cong 2.06$, $E \cong 152\,MeV$) but with bunch lengths corresponding to few (about 3 and 6) coherence lengths. The simulation shows a sharp dependence of the saturated power on $\alpha_\varepsilon$ which exhibits a maximum for negative value. This is just one aspect of a fairly interesting phenomenology, involving the FEL pulse dynamics and the relevant shape.

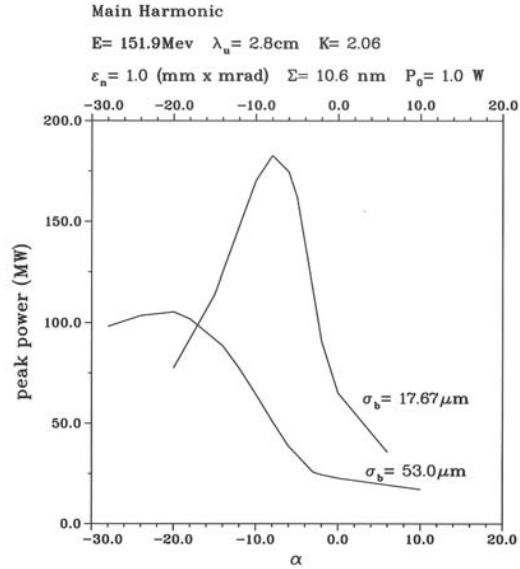

**Fig. 1**

**Peak power vs. $\alpha_\varepsilon$ for electron bunches with the same charge but different lengths a) $\sigma_z = 17.67\,\mu m$, peak current $I_e = 159\,A$, b) $\sigma_z = 53.0\,\mu m$, peak current $I_e = 53\,A$, Σ=10.60 nm, $\sigma_\varepsilon \cong 6 \cdot 10^{-4}$**

In this paper we discuss these effects and the relevant consequences on the harmonic generation and we will see how the energy correlation parameter may become a key quantity in the control of the FEL output beam quality.

# 2 Pulse shape and non linear Harmonic Generation

The FEL SASE operation with a beam exhibiting an energy phase correlation not only affects the maximum power, but also the laser pulse dynamics and shapes.

This is evident from fig. 2 where we have reported the evolution of the pulses in the region around the saturation point (before and after) for the cases with negative, positive and without correlation.

In the region above saturation the laser pulses exhibit the so called super-radiant behavior [7]. They develop typical side bands in their rear part, because it interacts with the electron bunch, thus gaining more energy than the front part, which tends to escape outside the electron bunch. The presence of the side bands is a combination of slippage and finite length of the electron bunch.

The side band growth is smoothened by a non vanishing correlation parameter, which controls the side band growth in a fairly efficient way. For negative values the pulse remains significantly narrower than the other two cases with a significantly larger peak of the power.

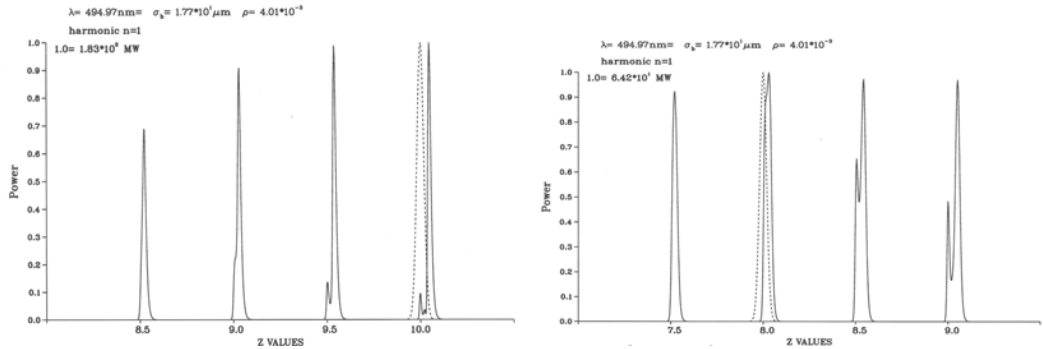

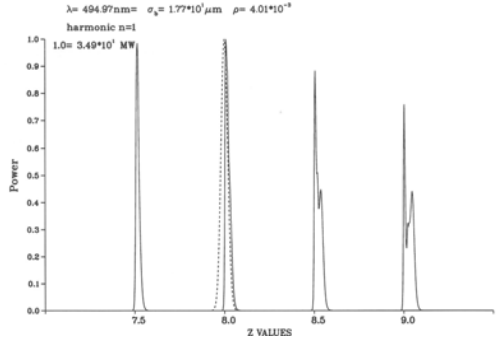

**Fig. 2**

**Pulse shape evolution vs. the undulator longitudinal coordinate for the case $\sigma_z = 17.67\,\mu m$, the dotted curve denotes the electron bunch distribution and specifies the position at which saturation (understood as the maximum power of the peak of the pulse) occurs.**

**a)** $\alpha_\varepsilon = -8$, **b)** $\alpha_\varepsilon = 0$, **c)** $\alpha_\varepsilon = 6$

The physical reasons underlying this behavior will be discussed later in section. Here we note that the control of the side-band growth is essentially due to the fact that the correlation parameter affects the slippage mechanism. For positive values, the lethargic effect, namely the slowing down of the laser pulse velocity due to the interaction and consequent gain, is enhanced. The electron and optical bunches overlaps for most of the time and therefore the rear and front part of the bunch experiences nearly equal gain factors. In absence of the correlation the slippage is not sufficiently counteracted by the lethargy and the side band grows. For negative values the center of mass move faster, it is pulled outside abruptly and the side band has not sufficient time to grow.

It is interesting to understand the consequence of the above dynamics on the non linear harmonic generation, which seems strongly enhanced for a beam with negative $\alpha_\varepsilon$ values (See Figs. 3).

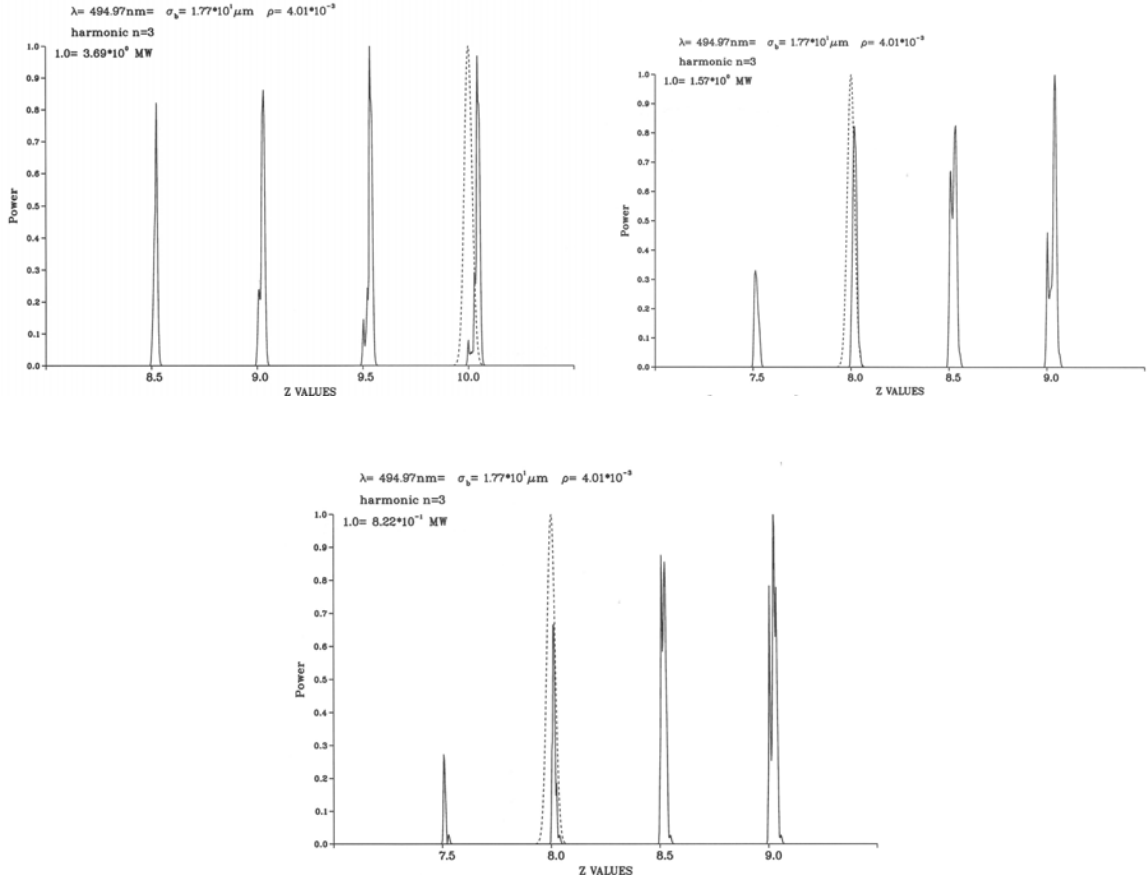

Fig. 3

Same as Fig. 2) for the third harmonics

**a)** $\alpha_\varepsilon = -8$, **b)** $\alpha_\varepsilon = 0$, **c)** $\alpha_\varepsilon = 6$

The physical reasons determining this effect are just due to the fact that the shorter laser pulse emerging in the operation with negative correlation parameter determines a more efficient bunching, since a quite robust pulse interacts with almost fresh electrons, because the interaction occurs essentially on the border of the trailing edge of the electron bunch.

The effects we have pointed out appear quite remarkable and are peculiar of either the correlation factor and the shortness of the electron bunch. It is however interesting to consider the laser pulse shape evolution for larger values of the electron bunch length.

In Fig. (1) we have reported the behavior of the output power on the correlation parameter for two different electron bunches, having the same charge but different lengths. The figure shows a significant, but less sharp, dependence on $\alpha_\varepsilon$. In particular, it appears more flat for larger negative $\alpha_\varepsilon$ values. The pulse shape evolution for the fundamental harmonic are shown in Figs. 4. The same comments as before hold for this case too, and we find indeed that for negative correlation values the pulse remains narrow and does not develop any side band and remains on the front edge of the electron bunch even for large values of $\sigma_z$.

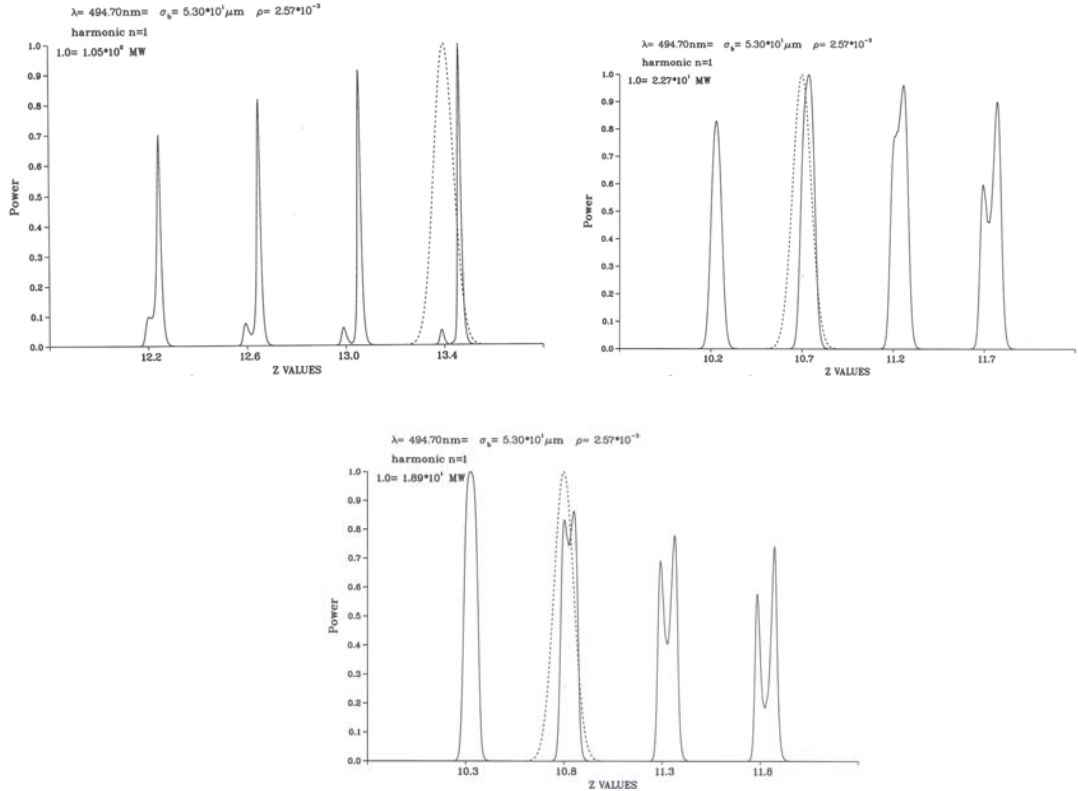

**Fig. 4**

**Same as Fig. 2 for $\sigma_z \cong 53 \mu m$**

**a)** $\alpha_\varepsilon = -10$, **b)** $\alpha_\varepsilon = 0$, **c)** $\alpha_\varepsilon = 6$

The structure of the pulse of the fundamental determines the quality of the third harmonic, as shown in fig. 5.

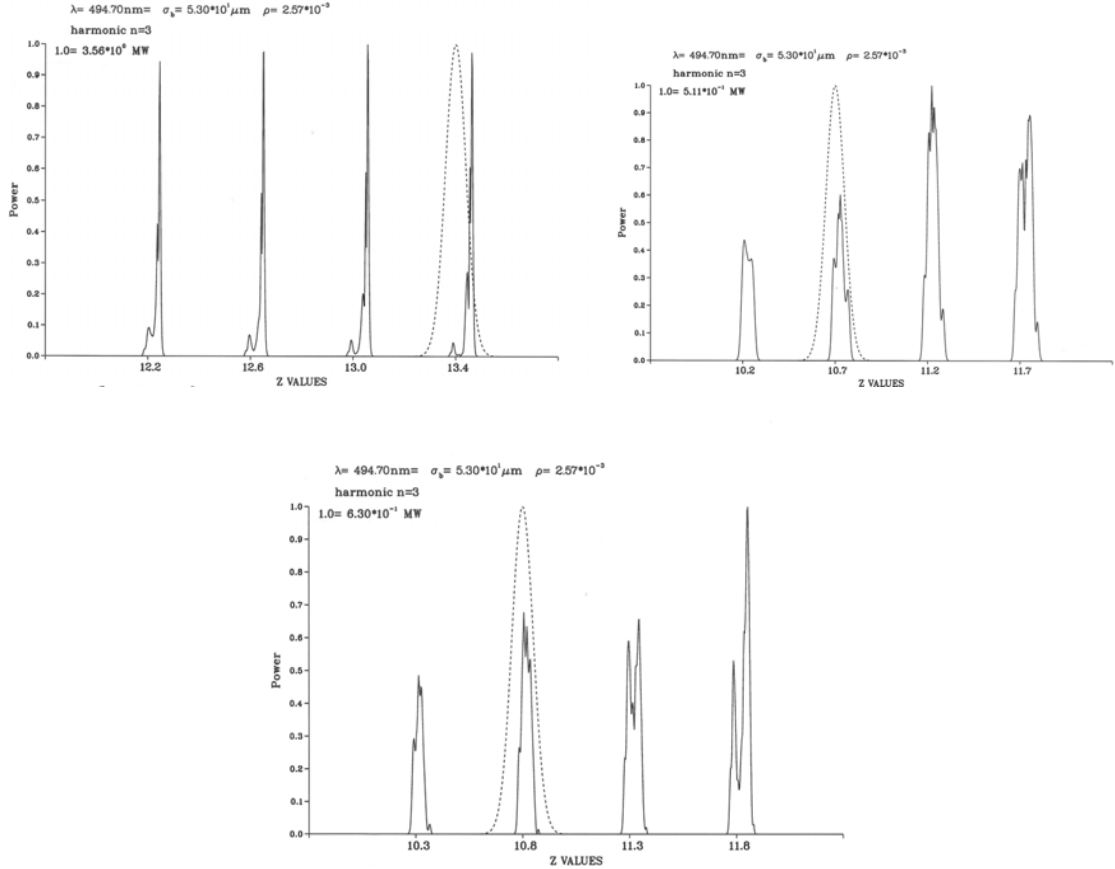

**Fig. 5**

**Same as Fig. 4 for the third harmonics**

**a)** $\alpha_\varepsilon = -10$, **b)** $\alpha_\varepsilon = 0$, **c)** $\alpha_\varepsilon = 6$

According to the results presented in this section the effect of the correlation parameter is responsible for a very interesting pulse dynamics, which certainly deserves careful considerations.

# 3 Concluding remarks

The phenomenology discussed in the previous sections can be explained on the basis of the following argument.

Consider the operation with negative correlation: the front part of the bunch has larger energy and emits radiation at shorter wave length, while the rear part emits radiation at longer wavelength. The radiation emitted in the front part starts to grow while that emitted on the tail is completely out of the gain curve. In these conditions only the trailing edge of the electron bunch is active. The shortening of the pulse is determined by the fact that the pulse has a very narrow region where it can gain.

For positive values the situation is just reversed, the back part of the beam is radiating at shorter wavelength and this ensures that, near saturation, this is the part of the beam emitting radiation within the gain bandwidth region.

We have already mentioned that the situation is reminiscent of some kind of tapering mechanism, this point has also been stressed in ref. [6]. We can therefore ask what happens if we combine tapering and energy correlation. In fig. 6 we report the result of a simulation including the energy correlation and undulator tapering, which is the result of a one dimensional optimization for the uncorrelated case.

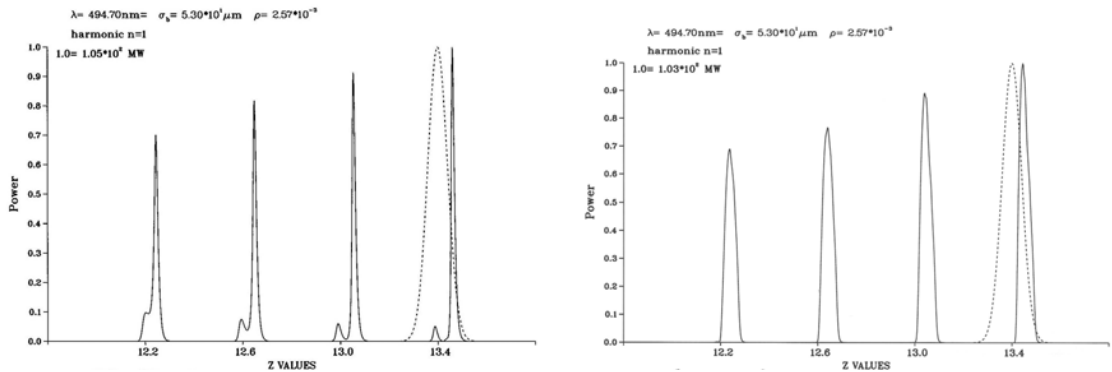

**Fig. 6**

**Pulse propagation with energy correlation and tapering. Same parameters of Fig. 2** $\sigma_z \cong 53 \mu m$ **a)** $\alpha_\varepsilon = -20$ **b)** $\alpha_\varepsilon = 8$

The results shown in the figure can be explained as it follows. The undulator tapering is in some sense counteracting that of the correlation. For negative correlation we must increase (in absolute value) $\alpha_\varepsilon$ to get the same effect of the untapered case. For positive values the combined effect of correlation and tapering is that of eliminating the growth of the side bands.

The modification induced by the correlation term in the high gain FEL small signal equation can be included in the FEL small signal high gain equation as it follows

$$\partial_\tau a(z,\tau) = i\pi g_0 \int_0^\tau e^{-i\nu\tau' - \frac{(\pi\pi_\varepsilon\tau')^2}{2}} \Phi(z+\Delta\tau,\tau') a(z+\Delta\tau',\tau-\tau') d\tau',$$

$$\Phi(z,\tau) = \exp\left\{-\frac{1}{2}\left[\left(\frac{z}{\sigma_z} + i\pi\mu_{\varepsilon,c}\tau\right)^2 + \mu_{\varepsilon,c}^2\tau^2\right]\right\}, \quad (5)$$

$$\mu_{\varepsilon,c} = 2N\alpha_\varepsilon, \quad \mu_\varepsilon = 4N\sigma_\varepsilon$$

where $a(z,\tau)$ is the Colson' dimensionless FEL amplitude, $g_0$ is the small signal gain coefficient, $\nu$ the frequency detuning parameter, $\tau = \frac{ct}{L_u}, L_u = N\lambda_u$ and $N$ is the number of the undulator periods. In deriving eq. (5) we have assumed that the electron bunch has a Gaussian shape

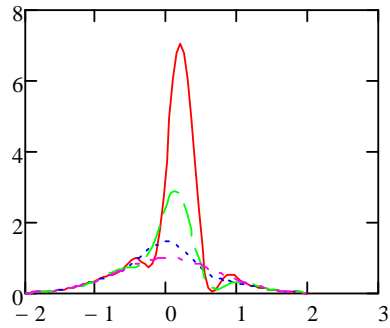

**Fig. 7**

**Growth of the FEL pulse at different times inside the undulator**

Eq. (5) accounts pulse propagation effects and high gain, but far from the saturation, it does not include indeed any non linear contribution in the field amplitude, which can be inserted in a phenomenological way to model the saturated dynamics using a kind of Ginzburg Landau model.

Postponing this procedure to a forthcoming investigation, we note that one of the main modifications with respect to the uncorrelated case is that the detuning parameter should be modified as

$$v \to v + \frac{\pi \alpha_{c,\varepsilon}}{\sigma_z} z \qquad (6)$$

which yields that, for negative values of the energy correlation parameter, the gain is larger for z position in the front of the bunch.

This is also confirmed by the numerical integration of eq. (5) reported in Fig. 7, which substantially agrees with the ab-initio numerical computation reported in the previous sections.

According to eq. (4) we can also conclude that the relative energy shift, between different positions in the bunch, produces a kind of line inhomogeneous broadening, because, at each position in $z$, corresponds a slightly different wave length. The associated relative line broadening is easily calculated for a Gaussian beam distribution and we get

$$\left\langle \frac{\delta \lambda}{\lambda} \right\rangle = 2 \frac{\alpha_\varepsilon}{\beta_\varepsilon} \sigma_z \qquad (7)$$

The importance of the effects induced by the above line broadening, can be guessed by comparing (7) to the FEL high gain bandwidth, namely

$$\tilde{\mu}_\alpha = 2 \frac{\alpha_\varepsilon}{\beta_\varepsilon} \frac{\sigma_z}{\rho} \qquad (8).$$

The impact of energy correlation on the FEL dynamics can be neglected, whenever the following inequality is fulfilled

$$\sigma_z \ll \frac{1}{2} \frac{\beta_\varepsilon}{\alpha_\varepsilon} \rho \qquad (9).$$

The above condition is more difficult to be satisfied for large correlation coefficient and for small FEL gain parameter.

**ACKNOWLEDGMENTS**

**The Authors express their sincere appreciation for stimulating discussions to Drs. M. Quattromini and D. Filippetto.**